\documentclass{article}
\usepackage{spconf,amsmath,graphicx}
\usepackage{hyperref}

\title{Deep Learning Object Detection Approaches to Signal Identification}
\name{Luke Wood$^{1,2}$, Kevin Anderson$^1$, Peter Gerstoft$^1$, Richard Bell$^1$, Raghav Subbaraman$^1$, and Dinesh Bharadia$^1$
\thanks{Our fully open source codebase is available at \url{https://github.com/lukewood/em-loader}
}}
\address{$^1$ University of California, San Diego, La Jolla, CA, USA \\
$^2$  Google, San Diego, CA, USA}
\begin{document}
\maketitle

\begin{abstract}
Traditionally Electromagnetic signal identification is  using threshold-based energy detection algorithms.
These algorithms frequently sum up the activity in regions, and consider regions above a
specific activity threshold to be signal.
While these algorithms work for the majority of cases, they often fail to detect signals
that occupy small frequency bands, fail to distinguish signals with overlapping frequency
bands, and cannot detect signals below a specified signal-to-noise ratio.
Through the conversion of raw signal data to spectrogram, source identification can be
framed as an object detection problem.
By leveraging advancements in deep learning based object detection,
we propose a system that manages to alleviate the failure cases encountered when using
traditional source identification algorithms.
Our contributions include framing signal identification as an object detection problem,
the publication of a spectrogram object detection dataset, and evaluation of the RetinaNet and
YOLOv5 object detection models trained on the dataset.
Our models achieve Mean average Precisions (MaP) of up to 0.906.  With such a high MaP,
these models are sufficiently robust for use in real-world applications.
\end{abstract}

\section{Introduction}

\begin{figure}
    \centering
    \includegraphics[width=0.75\linewidth]{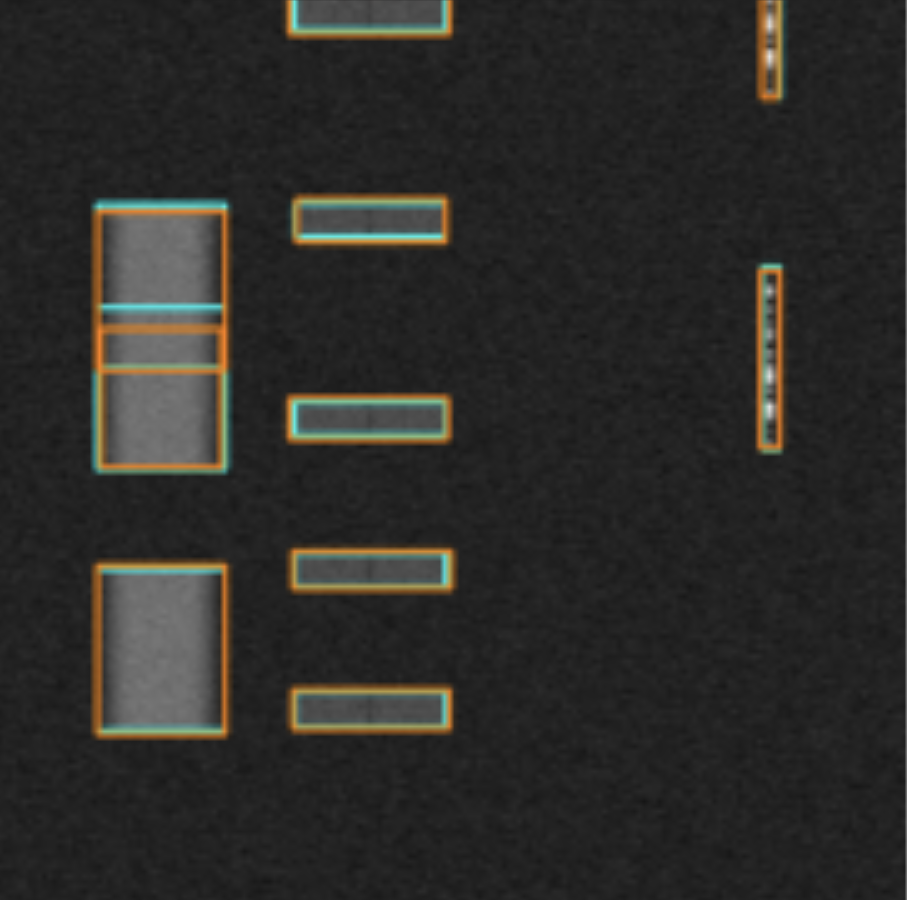}
  %  \caption{Predicted bounding boxes from a sampling of images from the validation set using the YOLOv5 Model}
    \label{fig:yolov5_predictions}
%\end{figure}
%\begin{figure}
    \centering
   \includegraphics[width=0.75\linewidth]{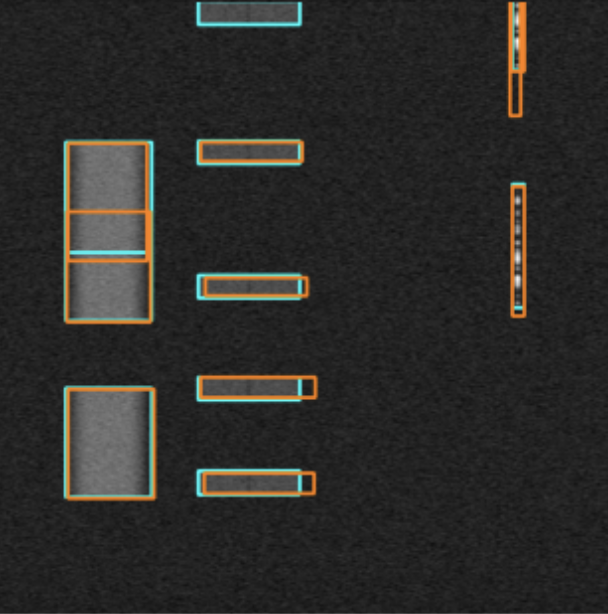}
%     \caption{Predicted bounding boxes from a sampling of images from the validation set using the RetinaNet model}
    \caption{A sampling of detections (orange box) and ground truths (cyan box) using the YOLOv5 (top) RetinaNet (bottom) model.  Background spectrum grey-scale is normalized form 0-1.
    The frequency spans 100 MHz on the x-axis is  and the time spans 50 ms on the y-axis. 
    Notice the overlapping signals to the left.}
    \label{fig:ground-truth}
\end{figure}

\label{sec:intro}
%{\color{red} Peter can you help us write about signal identification?}
%{\color{red} Cut down to 4 pages}

Since the inception of YOLO\cite{yolo}, the object detection space has been dominated by deep learning based architectures.
With the popularization of deep learning based object detection approaches, object
detection models have become accuracy, robust and reliable.  

By reframing signal identification as an object detection problem, we translate these
advancements in deep learning based object detection to the signal detection problem.
Our system  tackle the failures cases of traditional activity-threshold based
systems.
These failure cases include failing to distinguish distinct signals, failure to detect
signals occupying small frequency bands, and failure to detect signals below the signal
to noise floor.  
One such case where traditional models fail is demonstrated in figure \ref{fig:ground-truth}.  In the left portion of the sample spectrogram multiple signals occupy a shared frequency space.  This causes energy based detectors to identify them as a single signal.  Our deep learning object detection approach to signal identification successfully distinguishes and detects both signals.

Alongside our publication we open source everything required to reproduce our
results.
This includes our annotated spectrogram dataset, training pipeline, and inference
pipeline.  Our final models include both YOLOv5 and RetinaNet models, capable of scoring 0.906 Mean Average Precision (MaP) on the dataset.
Figure \ref{fig:ground-truth} shows nine frames annotated ground truth samples (blue) from
the generated EM spectrum dataset along with detected bounding boxes (red). 
In a real-time system the frequency axis (x-axis) is 100 MHz and time axis (y-axis) is 50 ms each frame. Thus, we need to process 20 frames per second. % Peter please verify the validity of this claim

Our contributions include the framing of signal identification as an object detection
problem, an object detection dataset consisting of spectrograms and
bounding box annotations to accompany, a deep learning based signal identification
system trained on the novel dataset, analysis of the distinctions between traditional
object detection datasets and our dataset, and finally a comparison
between our deep learning based approach and traditional energy based systems.
The contributions can be broken into three components: problem framing,
dataset production, and model training.

%\section{Related Works}
We present a novel object detection dataset as well as propose solutions to the dataset
including a trained RetinaNet\cite{retinanet} and YOLOv5\cite{glenn_jocher_2022_7002879}
architecture.
YOLOv2 \cite{oshea2017} and YOLOv3 has been used for EM spectrum detection obtaining 85\% average prediction in their dataset  \cite{vagollari2021joint}.

Datasets of similar format include Common Objects in Context\cite{lin2014microsoft},
Pascal VOC\cite{everingham2010pascal}, and Kitti\cite{geiger2012we}.
Our dataset differs primarily in the fact that instead of natural images, it consists of
spectrogram images of EM signals.

Other object detection architectures that could be used to solve our dataset include
Faster-RCNN \cite{ren2015faster}, DINO SWIN-L\cite{liu2021swin},
YoloX\cite{ge2021yolox}, and Yolov3\cite{redmon2018yolov3}.

We evaluate our models using MSCOCO metrics.  These metrics were originally explained in
the Common Objects in Context publication\cite{lin2014microsoft}.  To evaluate our
MSCOCO metrics we use both the pycocotools and KerasCV metric implementations
\cite{wood2022kerascv}.  

While the COCO variant of Recall measures the models ability to detect all of the boxes in the validation set, the COCO variant of MaP incentivizes the detection of all boxes in the validation set  while also punishing false positive.  Mean Average Precision can be interpreted as a general purpose metric for the evaluation of an object detection model.  More information on metric definitions and the process of their computation can be found in Ref.\ \cite{cocometrics}.
%\textit{Efficient Graph-Friendly COCO Metric Computation for Train-Time Model Evaluation}\cite{cocometrics}.

\section{Data Generation}
\label{sec:datagen}
To demonstrate the effectiveness of our methodology, we present a synthetic dataset generated by Matlabs Communications Toolbox.
Our dataset consists of 4686 annotated spectrogram samples.
These samples contain a 512x512 image and 0-13 signals, each with a corresponding bounding box annotation.

The samples in the dataset consists of a spectrogram containing a variety of signals
and annotations of the signal bounding boxes.
All spectrograms were constructed from I/Q samples that were generated using the Matlab
communications and signal processing toolboxes with a sampling rate of 100 MHz and a
total transmission time of 50 ms.  % Kevin please address what MS is
The signal types appearing in this dataset are direct-sequence spread spectrum (DSSS), Bluetooth low energy (BLE), quadrature amplitude modulation (QAM), amplitude modulation (AM), frequency modulation (FM), and WiFi with
each sample having between one and four signals from this set.
Signal metadata like center frequency, bandwidth, arrival time, and
signal-to-noise ratio (SNR) are uniformly randomized between samples.
%{\bf Please give details}

To generate samples for the dataset, a combination of signal types is first selected. Next, we initialize the signal and add white Gaussian noise. The center frequency, bandwidth, and SNR for each signal type are randomly selected, a signal generator is instantiated, and each signal is added to the source. After this signal metadata is configured, I/Q samples for 5 realizations of this configuration are generated with signal durations and arrival times being randomized between realizations.

We repeat this process of randomly selecting signal metadata and generating realizations 20 times for a given combination of signals, resulting in 100 total realizations of signal-level metadata for every combination. After all realizations are generated, the source is cleared, the I/Q samples and metadata are stored, and the entire process is repeated for a new combination of signals.

Spectrograms are constructed from the I/Q samples using a sampling rate of 100 MHz,
1024 discrete Fourier transforms, an overlap length of 128 samples. %and a Hanning window of length 256. % Kevin please clarify meaning of the hanning window.
The spectrograms are resized to 512x512 with bicubic interpolation and saved as PNG images. The coordinates for the bounding box annotations for the signals in an image are calculated from the signal metadata and saved in a corresponding text file. These images and labels are the dataset used in training the models.

\subsection{Difficulties of Spectogram Object Detection}
\label{sec:anchor_configuration}
In our initial experiment, we train a RetinaNet with the library default settings for
the RetinaNet.  This includes the default settings for IoU threshold for label encoding,
IoU threshold for label decoding, and the default configuration for the AnchorBox
generation process.

In  initial experiments, the loss converged to low values on both the
training and validation sets.
Despite this, the model only achieves an MaP of
0.205 and a Recall of 0.238.
These are intolerable results for use in a production system.

%Upon investigation of some sample predictions, it becomes clear that the model is able to detect some
%source classes with both high precision and recall, while entirely missing other source
%classes.
%The medium sized signals with roughly equal side lengths are detected by the
%detector with low variance between the ground truths and the predicted boxes.
%The small
%boxes,  those with high aspect ratios, those with low aspect ratios, and large boxes are
%never detected at all.

The reason that the loss converges to such a low loss but the MaP and Recall remain low
stems from the regular shapes of objects in the EM spectrograms.
While most objects in the natural world follow a smooth distribution for aspect ratio  spectrograms have no such property.
This leads the signals present in the spectrograms to have a wide array of aspect ratios.
As such, the anchor boxes generated by the default configuration used in the KerasCV
RetinaNet are unlikely to match with the boxes from our training dataset in the label
encoding process.
This leads to the boxes that are not encoded having no representation in the encoded
batch of encoded training targets.
This explains the low loss, low Recall, and low MaP.

\begin{figure}
    \centering
    \includegraphics[width=0.49\linewidth]{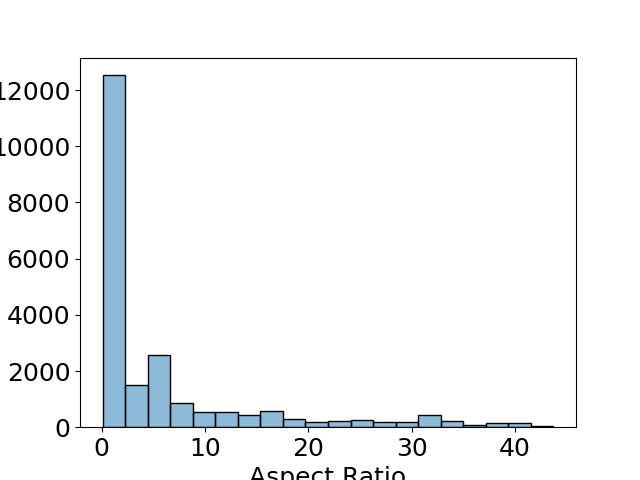}
    \includegraphics[width=0.49\linewidth]{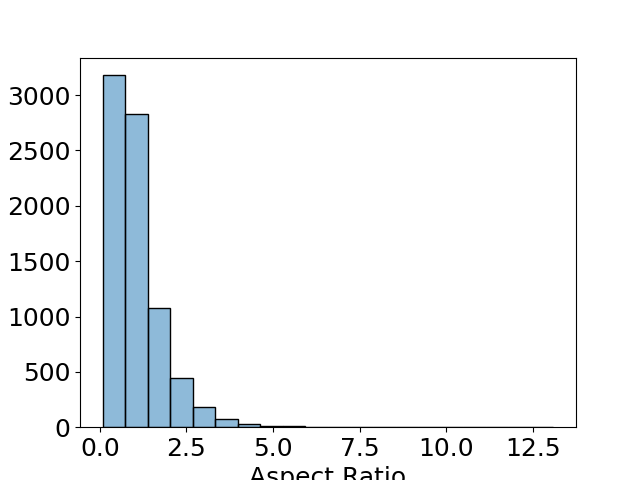}

    \caption{Aspect ratio histogram of spectrograms (left) and PascalVOC objects (right).}
    \label{fig:aspect_ratios}
\end{figure}

\begin{figure}
    \centering
    \includegraphics[width=0.49\linewidth]{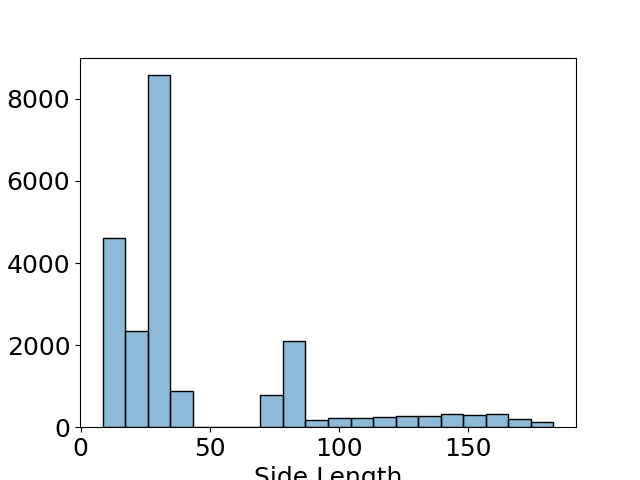}
    \includegraphics[width=0.49\linewidth]{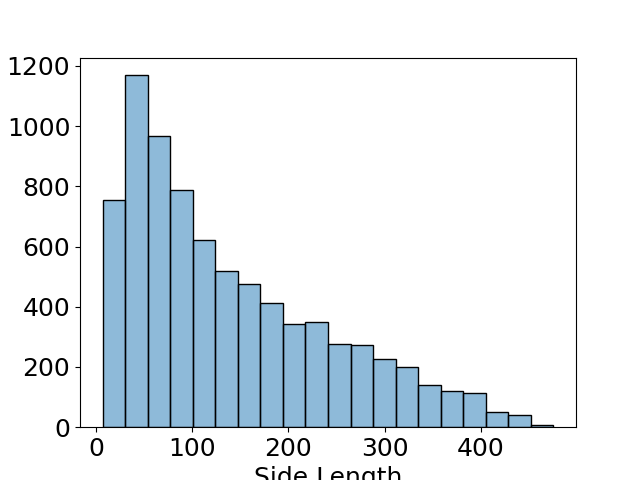}
    \caption{Side length in pixels histogram of spectrograms (left) and. PascalVOC objects (right).}
    \label{fig:side_lengths}
\end{figure}

The reason that anchor configuration is particularly important for spectrogram object
detection is attributed to the wide,  distribution of aspect ratios and side lengths of the spatial
representations of the bounding box annotations.
This is shown in a histograms in
figures \ref{fig:aspect_ratios} and \ref{fig:side_lengths}.

This differs greatly from object annotations present in natural images.
For comparison, the aspect ratio and side
length distributions for PascalVOC can be seen in figures \ref{fig:aspect_ratios} and \ref{fig:side_lengths}.
%These distributions are much closer to that of a Gaussian distribution.
    
\section{Training \& Experimental Results}
\label{sec:experimental_results}

Using the dataset produced using the process described in section \ref{sec:datagen},
we can train any deep learning based object detection model.
Our models are trained on a pre-generated train test split.
This split consists of 3859 training images, and 827 test images.
The object detection models are evaluated based on two metrics: the standard variant of
MaP and Recall used in the MSCOCO challenge.
We leverage the KerasCV COCO metric, and parameterize them as described
in \cite{cocometrics}.
Using these implementations enables to perform train time evaluation of these metrics.

Due to the high cost of computation required to compute MaP
we do not evaluate the true MaP  during training.
Instead, the approximate  COCO metrics is obtained by evaluating for a subset of
20  images from the evaluation dataset {\bf define}.  Using this proxy, we 
evaluate our model's performance across epochs and monitor its train time progress.
Final metrics are evaluated on the entirety of the test set.
As an additional inference test, we manually examine the visual results of the predictions.

\subsection{RetinaNet}
\label{sec:training}
In the first experiment, RetinaNet is trained using the KerasCV\cite{wood2022kerascv} library.  The
RetinaNet\cite{retinanet} architecture uses a ResNet50\cite{resnet50v2} backbone, which 
achieve about 30 frames per second on a standard consumer GPU.  
With a MobileNetv3\cite{mobilenetv3} backbone, a RetinaNet can achieve up to 60 FPS on a consumer GPU, enabling real time signal identification.
We configure the aspect ratios for the anchor box generator according to the results of section \ref{sec:anchor_configuration}.
We do not configure the anchor generator according to the side lengths.  
This results in the model not detecting the small boxes.

During training, spectrograms are loaded into memory as tensors of shape $(512, 512, 3)$
 with pixel values in the range of $[0, 255]$.  Pixel values are rescaled to the
 range $[0, 1]$ by dividing  by 255.

The backbone used is a ResNet50, with weights initialized using the weights produced
by training a ResNet50 to perform image classification.
Our feature pyramid, prediction heads, and backbone are all trained using the
spectogram dataset.
A SGD optimizer with a global clip norm\cite{clipnorm} of $10.0$ is used for
fitting, with a batch size  8.
Without the global clip norm, the loss explodes due to steep gradients existing 
at many points in the loss landscape.
Training is lightweight; ours being performed on a single GPU A100.
The learning rate of our optimizer is reduced after 5 epochs of training with no loss
improvement on the validation set, and training stops once no improvement has been
seen in 20 epochs.

No data augmentation was used to train the RetinaNet model as it decreased performance.
  This is due to  that spectrograms have  different bounding box distribution than natural images.  As such, traditional
augmentation techniques such as RandomFlip, RandomShear, and others are not useful.

\begin{figure}
    \centering
    \includegraphics[width=0.475\linewidth]{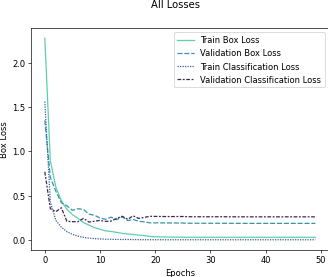}
    \includegraphics[width=0.475\linewidth]{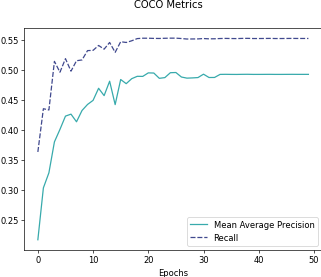}
    \caption{Losses (left) and COCO metrics (right) for the RetinaNet model}
    \label{fig:retinanet_metrics_losses}
\end{figure}

Losses alongside MaP and Recall metrics
are shown in figures \ref{fig:retinanet_metrics_losses}.  Results for
all metrics are  in Table \ref{table:final_metric_comparison}.
Upon convergence the RetinaNet model achieves a Recall of
0.565 and MaP of
0.492.

\subsection{YOLOv5}
\label{sec:yolov5}
To test the usefulness of various augmentation schemes and superior anchor configuration, we train a model using the pre-configured YOLOv5 framework maintained by Ultralytics\cite{glenn_jocher_2022_7002879}. The YOLOv5s architecture consists of a CSPDarknet53, a PANet feature pyramid, and a YOLOv4 head to generate the final output vectors with bounding boxes, class probabilities, and objectness scores.
The Ultralytics YOLOv5 framework automatically tunes anchor boxes for aspect ratio and side length, and includes the Mosaic augmentation\cite{glenn_jocher_2022_7002879} technique.  The mosaic augmentation is a logical choice for spectrogram object detection as due to the spatial meaning embedded in spectrograms translations, rotations, and color based augmentations all become meaningless.

A for RetinaNet, spectrograms are loaded into memory as tensors of shape (512,512,3) with normalized pixel values in the range [0,1]. A standard Keras stochastic gradient descent optimizer is used with a batch size of 16 and a warmup scheduler with a relatively low learning rate for 3 epochs as it ramps up to the normal learning rate. Training is done on a single NVIDIA T4 GPU over 50 epochs and set to stop once no improvement has been seen in 20 epochs.

\begin{figure}
    \centering
    \includegraphics[width=0.49\linewidth]{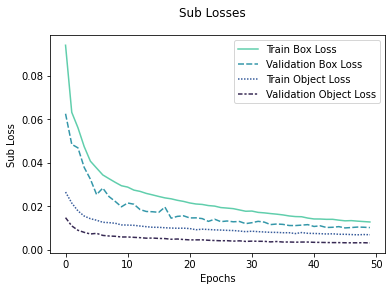}
    \includegraphics[width=0.49\linewidth]{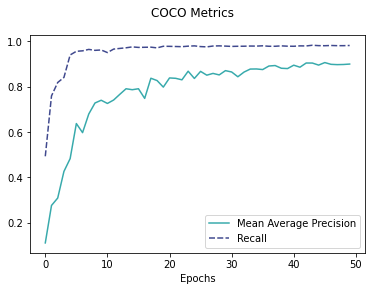}
    \caption{Losses (left) and COCO metrics (right) for the YOLOv5 Model}
    \label{fig:yolov5_all_losses}
\end{figure}

The loss function for YOLOv5 is a summation of a box loss, objectness loss, and classification loss. The classification loss is here ignored as there is only one class. Losses and COCO metrics are shown in Figure \ref{fig:yolov5_all_losses}.  The YOLOv5s model achieves much better  Recall and MaP than RetinaNet.

\subsection{Experimental Analysis}

Our experiments show that a deep learning based approach to signal identification is highly robust, the importance of anchor configuration in spectrogram object detection, and the effectiveness of the Mosaic data augmentation.
In training our RetinaNet, we initially did not tune the anchor generator at all.  
This yielded incredibly poor results, with a final MaP of  0.2.
Upon configuring only the aspect ratios, we scored an MaP of 0.492.  
Finally, we train a YOLOv5 model with optimal anchor configuration and achieve a MaP of 0.906.

In addition to the optimal anchor configuration, we determine that the mosaic augmentation significantly boosts performance.  This is a reasonable result; given that most other data augmentation techniques are no longer applicable in the spectral domain.  For example, rotations, color jitter, and many other common augmentations are
no longer suitable when working with spectrograms.  
Mosaic, on the other hand maintains the spatial structure of most the signals while still producing synthetic data.
Our two best models scores are available in table \ref{table:final_metric_comparison}.

\begin{table}[]
\begin{tabular}{l|ll}
\hline
          &  MaP & Recall    \\ \hline
YOLOv5    &  \textbf{0.906}                  & \textbf{0.980}     \\ 
RetinaNet &  0.492                  & 0.565     \\ \hline
\end{tabular}
\caption{Metrics for both the YOLOv5 and RetinaNet models.}
\label{table:final_metric_comparison}
\end{table}

\section{Conclusion}

Through reframing of signal identification as an object detection problem we leverage
advancements in deep learning based object detection to handle the failure cases of
traditional signal identification systems.
These failure cases include failing to distinguish distinct signals, failure to detect
signals occupying small frequency bands, and failure to detect signals below the signal
to noise floor.

We presented and open source a  dataset consisting of 7000 spectrograms alongside bounding box annotations
annotations of the signals present in these spectrograms, an open source Python library
to load the dataset into a TensorFlow dataset, an open source training script to train a
KerasCV RetinaNet on our novel dataset, and a sample solution to the dataset using YOLOv5.
We analyzed  the data to produce
optimal anchor box configuration for deep learning based object detection systems.
This show the distinct importance of anchor configuration in spectrogram object detection.

The YOLOv5 model achieved a  Mean average Precision (MaP) of
0.906 and Recall of 0.980.  
These metrics indicate that the model detects almost all signals boxes while making minimal false positive predictions.
Our model is sufficiently robust for deployment in a real world system.
\newpage
%\small
%\bibliographystyle{plain}
\bibliographystyle{IEEEbib}
\bibliography{abbrev1,refs}

\begin{thebibliography}{10}

\bibitem{yolo}
Joseph Redmon, Santosh Divvala, Ross Girshick, and Ali Farhadi,
\newblock ``You only look once: Unified, real-time object detection,'' 2015.

\bibitem{retinanet}
Tsung-Yi Lin, Priya Goyal, Ross Girshick, Kaiming He, and Piotr Doll{\'a}r,
\newblock ``Focal loss for dense object detection,''
\newblock in {\em Proceedings of the IEEE international conference on computer
  vision}, 2017, pp. 2980--2988.

\bibitem{glenn_jocher_2022_7002879}
Glenn~Jocher et~al.,
\newblock ``{ultralytics/yolov5: v6.2 - YOLOv5 Classification Models, Apple M1,
  Reproducibility, ClearML and Deci.ai integrations},'' Aug. 2022.

\bibitem{oshea2017}
Tim O'Shea, Tamohgna Roy, and T~Charles Clancy,
\newblock ``Learning robust general radio signal detection using computer
  vision methods,''
\newblock in {\em 2017 51st asilomar conference on signals, systems, and
  computers}. IEEE, 2017, pp. 829--832.

\bibitem{vagollari2021joint}
Adela Vagollari, Viktoria Schram, Wayan Wicke, Martin Hirschbeck, and Wolfgang
  Gerstacker,
\newblock ``Joint detection and classification of rf signals using deep
  learning,''
\newblock in {\em 2021 IEEE 93rd Vehicular Technology Conference
  (VTC2021-Spring)}. IEEE, 2021, pp. 1--7.

\bibitem{lin2014microsoft}
Tsung-Yi Lin, Michael Maire, Serge Belongie, James Hays, Pietro Perona, Deva
  Ramanan, Piotr Doll{\'a}r, and C~Lawrence Zitnick,
\newblock ``Microsoft coco: Common objects in context,''
\newblock in {\em European conference on computer vision}. Springer, 2014, pp.
  740--755.

\bibitem{everingham2010pascal}
Mark Everingham, Luc Van~Gool, Christopher~KI Williams, John Winn, and Andrew
  Zisserman,
\newblock ``The pascal visual object classes (voc) challenge,''
\newblock {\em International journal of computer vision}, vol. 88, no. 2, pp.
  303--338, 2010.

\bibitem{geiger2012we}
Andreas Geiger, Philip Lenz, and Raquel Urtasun,
\newblock ``Are we ready for autonomous driving? the kitti vision benchmark
  suite,''
\newblock in {\em 2012 IEEE conference on computer vision and pattern
  recognition}. IEEE, 2012, pp. 3354--3361.

\bibitem{ren2015faster}
Shaoqing Ren, Kaiming He, Ross Girshick, and Jian Sun,
\newblock ``Faster r-cnn: Towards real-time object detection with region
  proposal networks,''
\newblock {\em Advances in neural information processing systems}, vol. 28,
  2015.

\bibitem{liu2021swin}
Ze~Liu, Yutong Lin, Yue Cao, Han Hu, Yixuan Wei, Zheng Zhang, Stephen Lin, and
  Baining Guo,
\newblock ``Swin transformer: Hierarchical vision transformer using shifted
  windows,''
\newblock in {\em Proceedings of the IEEE/CVF International Conference on
  Computer Vision}, 2021, pp. 10012--10022.

\bibitem{ge2021yolox}
Zheng Ge, Songtao Liu, Feng Wang, Zeming Li, and Jian Sun,
\newblock ``Yolox: Exceeding yolo series in 2021,''
\newblock {\em arXiv preprint arXiv:2107.08430}, 2021.

\bibitem{redmon2018yolov3}
Joseph Redmon and Ali Farhadi,
\newblock ``Yolov3: An incremental improvement,''
\newblock {\em arXiv preprint arXiv:1804.02767}, 2018.

\bibitem{wood2022kerascv}
Luke Wood, Scott Zhu, Fran\c{c}ois Chollet, et~al.,
\newblock ``Keras cv,'' \url{https://github.com/keras-team/keras-cv}, 2022.

\bibitem{cocometrics}
Luke Wood and Francois Chollet,
\newblock ``Efficient graph-friendly coco metric computation for train-time
  model evaluation,'' 2022.

\bibitem{resnet50v2}
Kaiming He, Xiangyu Zhang, Shaoqing Ren, and Jian Sun,
\newblock ``Identity mappings in deep residual networks,'' 2016.

\bibitem{mobilenetv3}
Andrew Howard, Mark Sandler, Grace Chu, Liang-Chieh Chen, Bo~Chen, Mingxing
  Tan, Weijun Wang, Yukun Zhu, Ruoming Pang, Vijay Vasudevan, Quoc~V. Le, and
  Hartwig Adam,
\newblock ``Searching for mobilenetv3,'' 2019.

\bibitem{clipnorm}
Jingzhao Zhang, Tianxing He, Suvrit Sra, and Ali Jadbabaie,
\newblock ``Why gradient clipping accelerates training: A theoretical
  justification for adaptivity,'' 2019.

\end{thebibliography}

\end{document}